\documentclass[12pt,letterpaper,copyedit]{article}
\usepackage{osajnl2}
\usepackage[tablesfirst,notablist,nomarkers]{endfloat}

\usepackage{amsmath}
\usepackage{amsfonts}
\def\R{\mathbb{R}}

\begin{document}

\title{Generalized Cloaking and Optical Polyjuice}
\author{Andr\'e Nicolet,$^{1,*}$ Fr\'ed\'eric Zolla,$^1$ and Christophe Geuzaine$^{2}$}
\address{$^1$Institut Fresnel UMR CNRS 6133, Aix-Marseille Universit\'e, Ecole Centrale de Marseille, Campus de Saint-J\'er\^ome,  F-13013 Marseille, France}
\address{$^2$University of Li{\`e}ge, Dept. of Electrical
Engineering and Computer Science (ACE),
B-4000 Li{\`e}ge, Belgium}
\address{$^*$Corresponding author: andre.nicolet@fresnel.fr}


\begin{abstract}In this paper, a generalization of cloaking is presented: instead of an empty region of space, an inhomogeneous structure is transformed via Pendry's map in order to give, to any object hidden in the central hole of the cloak, a completely arbitrary appearance.
\end{abstract}

\ocis{000.4895, 260.2110}




In 2006, it was suggested by Pendry et al. \cite{pendrycloak} that an
object surrounded by a coating consisting of an exotic
material becomes invisible to electromagnetic
waves. This device was named ``invisibility cloak'' in reference to Harry Potter, the popular character of J.K. Rowling. Beside his famous cloak, the little wizard has other spells to go unnoticed. Among the most spectacular is the ``polyjuice potion'' that is able to turn somebody into anybody else's appearance \cite{jkrowling}. In this paper, we do not present a potion but rather an optical device able to accomplish the same task, i.e. to give an arbitrary optical response chosen in advance to any other object placed inside the device. In fact, the principle is here very similar to the design of Pendry's invisibility cloak but, instead of geometrically transforming an empty domain, it is a region containing the object to be imitated that is transformed leading thus to a generalization of cloaking.


In this section, we present a generalization of cloaking able to arbitrarily transform the electromagnetic appearance of an object. The basic principle is to obtain the constitutive relations of the cloak by application of a space transformation to a non-empty region.

A geometric transformation is given by a map $\varphi$ from a space $N$ to a space $M$. For all our practical purposes, $M$ and $N$ will be here the whole or parts of $\R^3$. Given a Cartesian coordinate system $\mathbf{x}$ on $M$ and an arbitrary coordinate system $\mathbf{x}'$ on $N$, $\varphi: N \rightarrow M$ is described by $\mathbf{x}(\mathbf{x'})$, i.e. $\mathbf{x}$ given as function of $\mathbf{x}'$. All the useful information, i.e. necessary to transform differential forms   (indeed the formalism of differential geometry is the most natural to write the Maxwell's equation \cite{deschamps,burke,nicolet,bossavit} specially when quite general geometric transformations are involved)   and other covariant tensors, is contained in the Jacobian matrix field $\mathbf{J}(\mathbf{x'})=\partial \mathbf{x}(\mathbf{x}')/\partial \mathbf{x}' $.

The basic principle of transformation optics is that, when you have an electromagnetic system described by the tensor fields $\underline{\underline{\varepsilon}}(\mathbf{x})$ for the dielectric permittivity and $\underline{\underline{\mu}}(\mathbf{x})$ for the magnetic permeability in the space $M$,  if you replace your initial material properties by equivalent material properties given by the following rule \cite{pcfbook,milton,twistedEPJ,twistedJWAves}:
      \begin{eqnarray}
      \underline{\underline{\varepsilon'}}(\mathbf{x'}) =
      \mathbf{J}^{-1}(\mathbf{x'})\underline{\underline{\varepsilon}}(\mathbf{x}(\mathbf{x'}))\mathbf{J}^{-T}(\mathbf{x'})\det(\mathbf{J}(\mathbf{x'})),  \notag \\
      \underline{\underline{\mu'}}(\mathbf{x'}) =
      \mathbf{J}^{-1}(\mathbf{x'})\underline{\underline{\mu}}(\mathbf{x}(\mathbf{x'}))\mathbf{J}^{-T}(\mathbf{x'})\det(\mathbf{J}(\mathbf{x'})),
      \label{equivalence_rule}
      \end{eqnarray}
 ($\mathbf{J}^{-T}$ is
the transposed of the inverse of $\mathbf{J}$), you get an equivalent problem on $N$. Here, an equivalent problem means that the solution of the new problem on $N$, i.e. electromagnetic quantities described as differential forms, are the pulled back of the solution \cite{nicolet} of the original problem on $M$ and that the same Maxwell's equations (i.e. as if we were in Cartesian coordinates or, more accurately, having the same form written with the exterior derivative) are still satisfied.

In the case of the cylindrical Pendry's map \cite{pendrycloak,opl,compel_geo}, described by the transformation of the 2D cross section, the plane $\R^2$ minus a disk $D_1$ of radius $R_1$ is mapped on the whole plane $\R^2$ in such a way that a disk $D_2$ of radius $R_2 > R_1$, concentric with $D_1$, is the image of the annulus $D_2\backslash D_1$ by a radial transformation. In cylindrical coordinates, this transformation is given by:

\begin{equation}\begin{cases}  r = (r'-R_1) R_2/(R_2-R_1) \; \mathrm{for} \; R_1 \leq r'\leq R_2, \; \\ \theta = \theta', \; z = z'. \;
\label{pendrysmap}\end{cases}\end{equation}

As for the outside of the disk $D_2$, the map between the two copies of $\R^2\backslash D_2$ is the identity map.

The material properties given by rule (\ref{equivalence_rule}) corresponding to this transformation provide an ideal invisibility cloak: outside $D_2$, everything behaves as if we were in free space, including the propagation of electromagnetic waves across the cloak, and is completely independent of the content of $D_1$.


Now, rule (\ref{equivalence_rule}) may be applied to $D_2$ containing objects with arbitrary electromagnetic properties so that a region cloaked by this device is still completely hidden but has the appearance of the objects originally in $D_2$. We may call this optical effect masking \cite{teixeira} or ``polyjuice'' effect.

\begin{figure}[h]
{\centering  \includegraphics[width=0.49\textwidth]{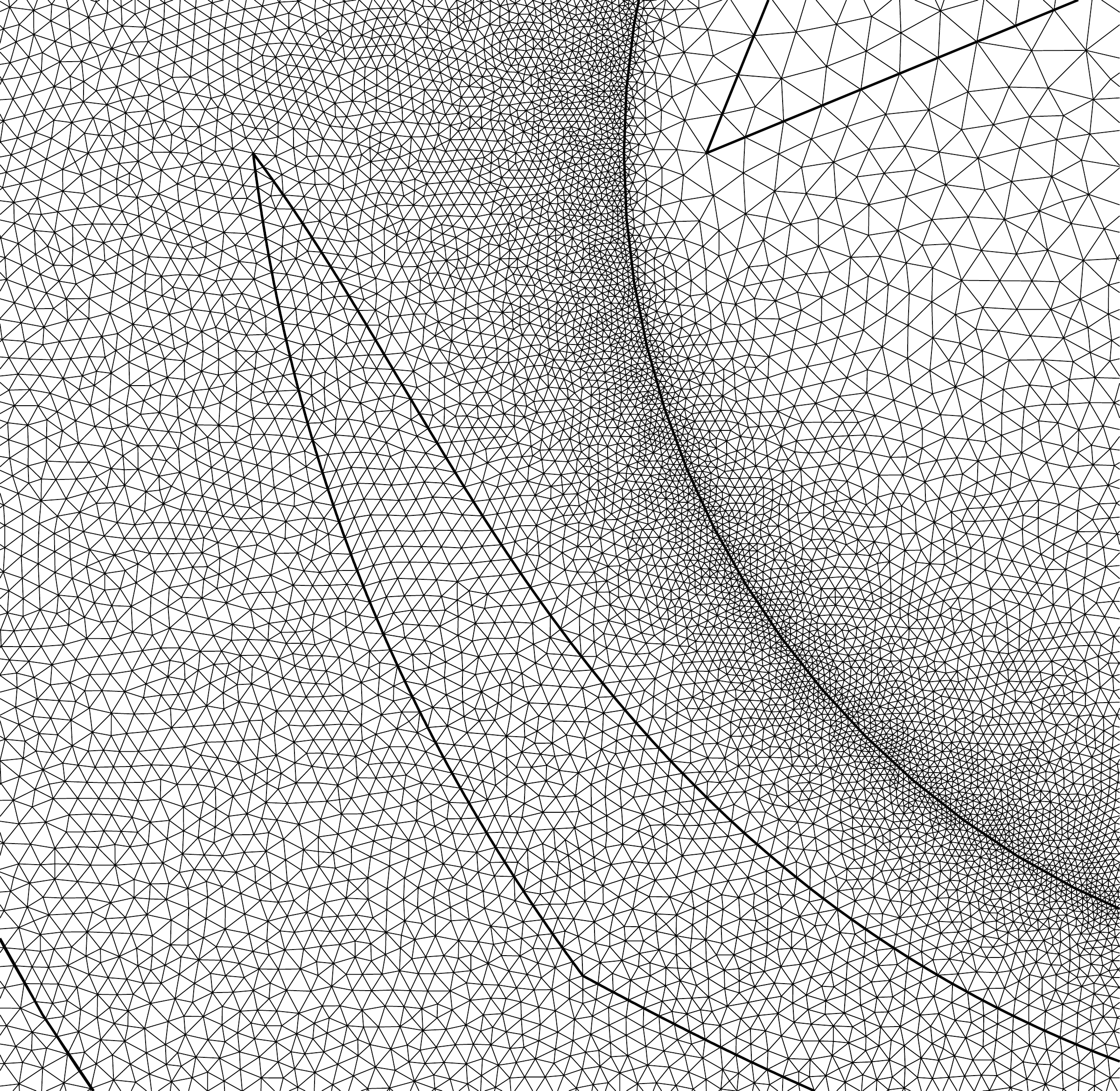}
\caption{\label{mesh}  This figure shows a part of the triangular mesh used for the finite element modeling of the scattering problem of  Fig. \ref{fig2}. The singular behavior of the permittivity and of the permeability requires a very fine mesh along the inner boundary of the cloak in order to achieve a satisfactory accuracy with the numerical model.  }}
\end{figure}

\begin{figure}[h]
{\centering \includegraphics[width=0.49\textwidth]{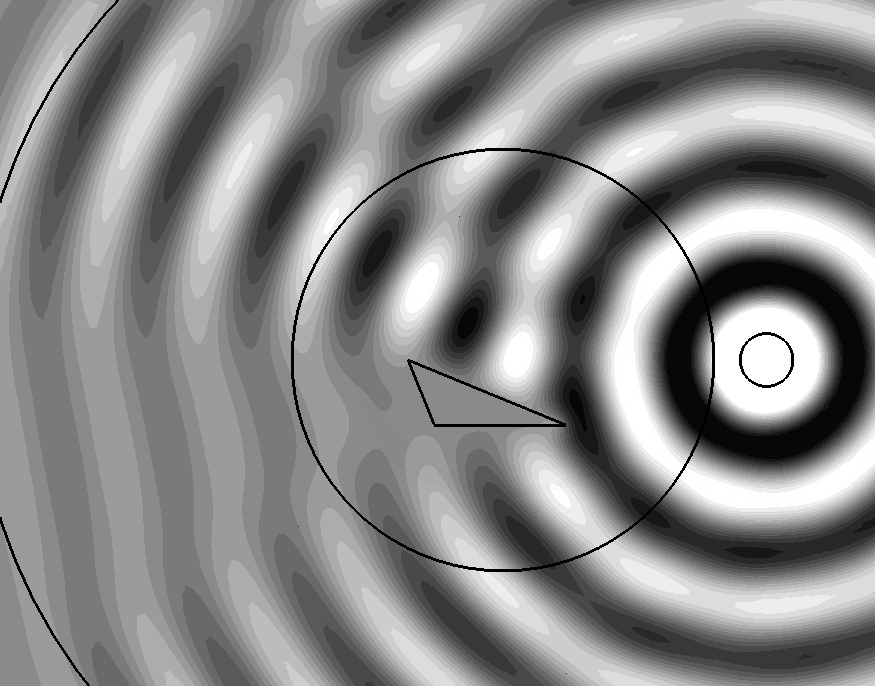}
\caption{\label{fig1}  A conducting triangular cylinder is scattering cylindrical waves.}}
\end{figure}

\begin{figure}[h]
{\centering  \includegraphics[width=0.49\textwidth]{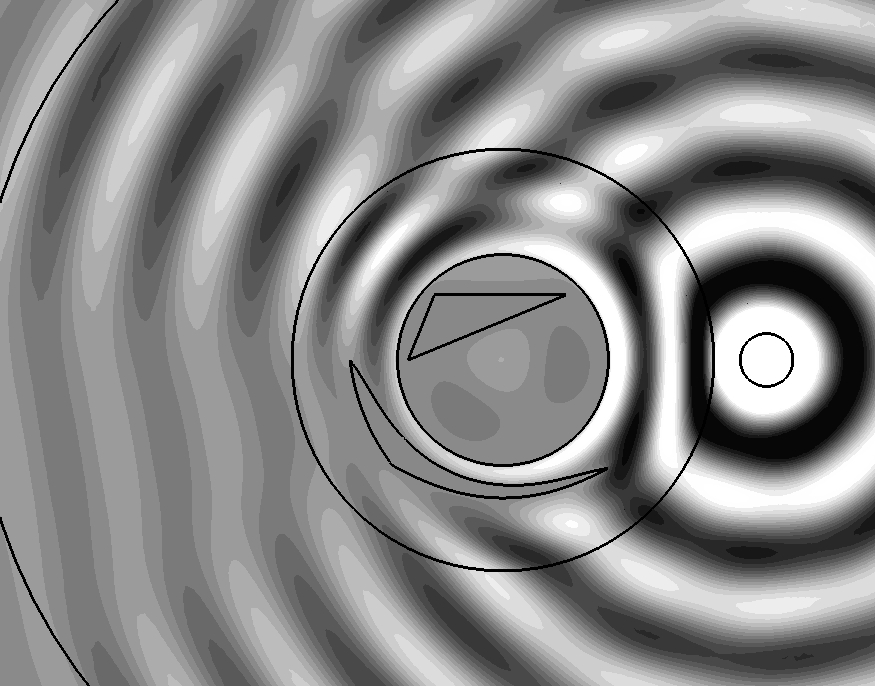}
\caption{\label{fig2}  A triangular cylinder different from the one on Fig. \ref{fig1} is surrounded by a cloak designed to reproduce the scattering pattern of the Fig. \ref{fig1} triangular cylinder in spite of the change of diffracting object. Of course, the diffracting object inside the cloak may be arbitrary as far as it is small enough to fit inside the cloak}}
\end{figure}

\begin{figure}[h]
{\centering  \includegraphics[width=0.49\textwidth]{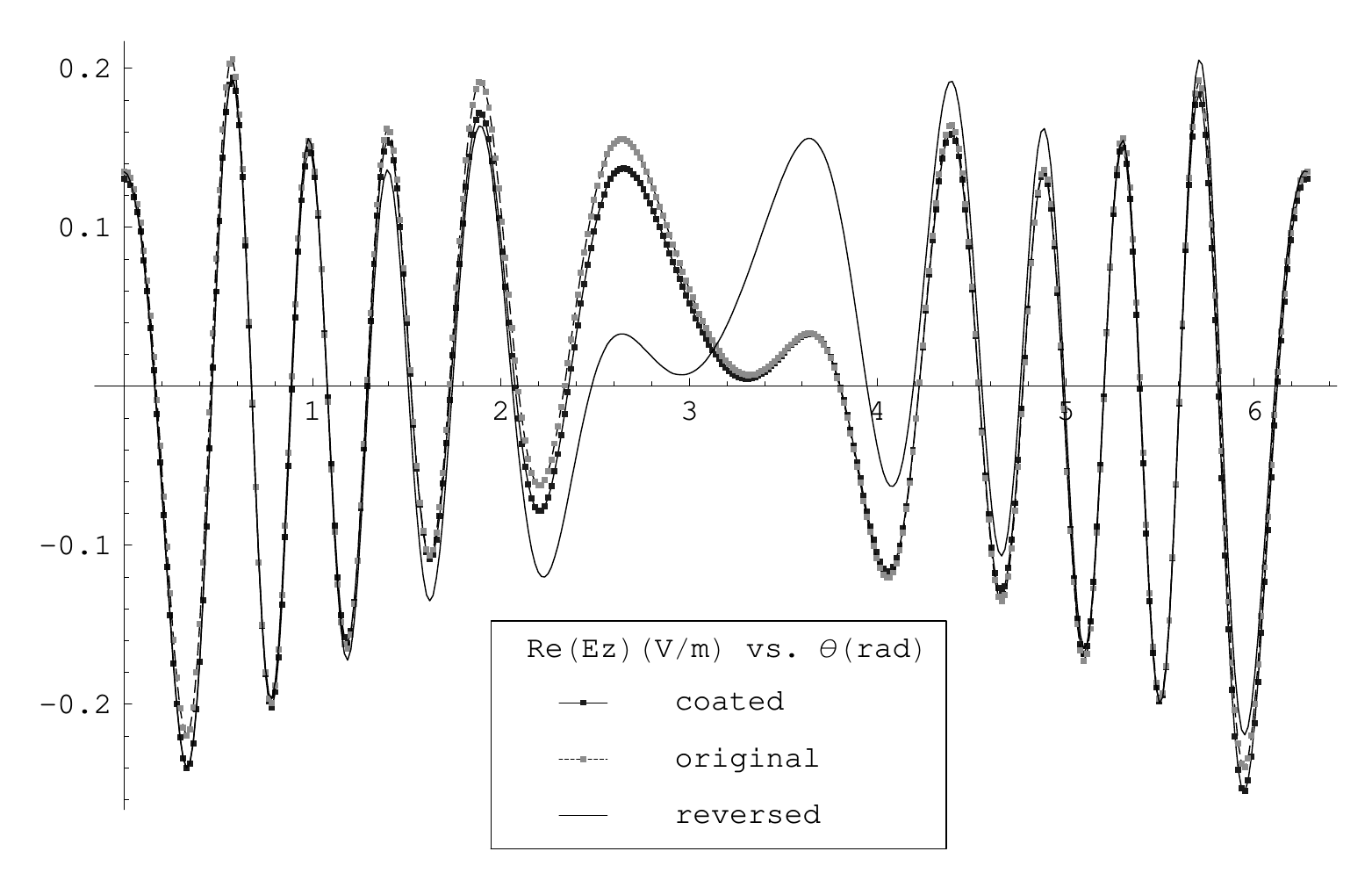}
\caption{\label{fig3}   The value of the electric field (the real part of $E_z$) on a circle of radius $4\lambda$ concentric with the cloak is represented as a function of the position angle $\theta$ (increasing counterclockwise and with $\theta=0$ corresponding to the point the most on the right). The three configurations considered here are the ones of Fig. \ref{fig2} (coated), Fig. \ref{fig1} (original), and the triangle of Fig. \ref{fig2} without the coating (reversed).  } }
\end{figure}


Figs. \ref{fig1} and \ref{fig2} show the effect of masking on a scattering structure. On Fig. \ref{fig1}, a cylindrical TM wave emitted by a circular cylindrical antenna is scattered by a conducting triangular cylinder (the longest side of the cross section is $1.62\lambda$ and $\varepsilon_r=1 + 40 i $). The field map  represents the longitudinal electric field $E_z(x,y)$ and the outer boundary of the cloak is shown to ease the comparison with the masked case. On Fig. \ref{fig2}, the same cylindrical TM wave is scattered by a masked triangular cylinder (but the diffracting object inside the cloak may be arbitrary as far as it is small enough to fit inside the cloak). This triangular cylinder is the symmetric of the previous one with respect to the horizontal plane containing the central fibre of the cylindrical antenna. This bare scatterer would therefore give the Fig. \ref{fig1} image inverted upside-down but, here, this object is surrounded by a cloak in order to give the very same scattering as before. Indeed, on both sides, the electric fields outside the cloak limit are alike.

  Fig. \ref{fig3} highlights the different scattering patterns by displaying the value of $\Re e(E_z)$ on a circle of radius $4 \lambda$ located around the antenna-scatterer system in the three following cases: the case of Fig. \ref{fig1} (original) with the triangle alone, the case of Fig. \ref{fig2} (coated), and the triangle of Fig. \ref{fig2} without the coating (reversed). It is obvious that the coating restores the field distribution independently of the object present in the central hole.

The numerical computation is performed using the finite element method (via the free GetDP \cite{getdp} and Gmsh \cite{gmsh} software tools). The mesh is made of 148,000 second order triangles   including the Perfectly Matched Layers used to truncate the computation domain . The singularity of $\varepsilon$ and $\mu$ requires a very fine mesh in the vicinity of the inner boundary of the cloak (see Fig. \ref{mesh}) and is also responsible for the small discrepancies between the numerical model and a perfect cloak (see Fig. \ref{fig2}) --- including the non zero field in the hole of the cloak.

Note that a small technical problem arises in practice when rule (\ref{equivalence_rule}) is applied: the material properties are defined piecewise on various domains and it is very useful to know explicitly the boundaries of these domains, e.g. to build the finite element mesh (see Fig. \ref{mesh}). These boundaries are curves in the cross section and are thus contravariant objects. Therefore, their transformation requires the inverse map $\varphi^{-1}$ from $M$ to $N$. Fortunately, map (\ref{pendrysmap}) is very simple to invert.

More explicitly, for a given curve $\mathbf{x}(t)$ of parameter $t$ in the initial Cartesian coordinates, its push forward by Pendry's map is:
\begin{equation}
\mathbf{x}'(t)=\varphi^{-1}(\mathbf{x}(t))=(\frac{R_2-R_1}{R_2}+ \frac{R_1}{\|\mathbf{x}(t) \|})\mathbf{x}(t),
\end{equation}
with the same variation of the parameter $t$. Note that the most common curves used in the design of devices, i.e. line segments and arc of circles, are transformed to less usual curves except   for radial segments (with respect to the center of the cloak) and arc of circles concentric with the cloak.

 On Fig. \ref{fig2}, the image by $\varphi^{-1}$ of the triangle of Fig. \ref{fig1} is the curvilinear triangle inside the coating region of the cloak. In practice, this anamorphosis of the triangle is described by three splines interpolating each 40 points that are images of points of the segments by $\varphi^{-1}$.


Transformation optics do not offer only the possibility make optically disappear objects in invisibility cloaks but also to completely tune their optical signature i.e. to give them an arbitrary appearance.
   The possibility to place an object inside the coating of the cloak and that it will therefore appear different was already considered in \cite{opl} where a point source was shifted creating a mirage effect. A more general case is considered here since both the shape and the position of the object placed in the coating are modified .
 Note that if the object used to create the illusion is perfectly conducting and surrounds the central point of the cloak, its anamorphosis will surround the inner boundary of the cloak and it \emph{de facto} suppresses the singular behavior of the material properties. In this particular case,   the present approach is similar to the the hiding under the carpet idea   of Li and Pendry \cite{carpet} but here a bounded object replaces the infinite reflecting plane. This technique can be naturally extended to cloaks of arbitrary shapes \cite{arbitrary}.


\begin{thebibliography}{1} 


\bibitem{pendrycloak} J.B. Pendry, D. Shurig, D.R. Smith,
"Controlling electromagnetic fields",
\newblock Science {\bf 312}, 1780 (2006).

\bibitem{jkrowling} J.K. Rowling, \emph{Harry Potter and the Chamber of Secrets}, (Bloomsbury Publishing PLC, 1998).


\bibitem{deschamps} G. A. Deschamps,"Electromagnetics and differential forms", \newblock Proc. IEEE,
\textbf{69}, 676 (1981).

\bibitem{burke} W. L. Burke, \emph{Applied Differential Geometry}, (Cambridge University Press, 1985).

\bibitem{nicolet} A. Nicolet, J. F. Remacle, B. Meys, A. Genon, W. Legros,
"Transformation methods in computational electromagnetism",
 J. Appl. Phys. {\bf 75}, 6036 (1994).

\bibitem{bossavit} A. Bossavit, J. Japan Soc. Appl. Electromagn.
\& Mech., \textbf{6}, "On the geometry of electromagnetism. (1): Euclidean space",17, "On the geometry of electromagnetism. (2): Geometrical objects",114, "On the geometry of electromagnetism. (3): Faraday's law",233, "On the geometry of electromagnetism. (4): 'Maxwell's house' ",318 (1998), \textbf{7}, "Computational electromagnetism and geometry: Building a finitedimensional
'Maxwell's house'. (1): Network equations",150, "Computational electromagnetism and geometry. (2): Network
constitutive laws",294, "Computational electromagnetism and geometry. (3): Convergence",401 (1999), \textbf{8}, "Computational electromagnetism and geometry. (4): From degrees of
freedom to fields",102, "Computational electromagnetism and geometry. (5): The 'Galerkin
hodge'",203, "Computational electromagnetism and geometry. (6): Some questions
and answers", 372 (2000).


\bibitem{pcfbook} F. Zolla, G. Renversez, A. Nicolet, B. Kuhlmey, S.
Guenneau, D. Felbacq, \newblock \emph{Foundations of Photonic Crystal Fibres},
(Imperial College Press, 2005).

\bibitem{milton} G. W. Milton, M. Briane, J. R. Willis,
"On cloaking for elasticity and physical equations with a transformation invariant form",
\newblock New J. of Phys.
\textbf{8}, 248 (2006).


\bibitem{twistedEPJ}   A. Nicolet, F. Zolla, S. Guenneau, "Modelling of twisted optical waveguides with edge elements", Eur. Phys. J. - Appl. Phys. {\bf 28}, 153 (2004).

\bibitem{twistedJWAves} A. Nicolet, F. Zolla, Y. Ould Agha, S. Guenneau, "Leaky modes in twisted microstructured optical fibers",  Waves in Random and Complex Media {\bf 17}, 559 (2007).


\bibitem{opl} F. Zolla, S. Guenneau, A. Nicolet, J. B. Pendry,
"Electromagnetic analysis of cylindrical invisibility cloaks and the mirage effect",
\newblock  Opt. Lett. \textbf{32}, 1069 (2007).

\bibitem{compel_geo}
A. Nicolet, F. Zolla, Y. Ould Agha, S.
Guenneau, "On the use of PML for the computation of leaky modes : an application to gradient index MOF",
\newblock COMPEL \textbf{27}, 806 (2008).

\bibitem{teixeira}
F. L. Teixeira, "Differential form approach to
the analysis of electromagnetic
cloaking and masking", Microwave Opt. Technol. Lett. {\bf 49},  2051 (2007).

\bibitem{getdp}
P. Dular, C. Geuzaine, F. Henrotte, W. Legros,
\newblock "A general environment for the treatment of discrete problems and its application to the finite element method",
 IEEE Trans. Mag.  { \bf 34},
3395 (1998). (see also http://www.geuz.org/getdp/)

\bibitem{gmsh}
C. Geuzaine, J.-F. Remacle,
"Gmsh: a three-dimensional finite element mesh generator with built-in pre- and post-processing facilities",
Int. J. Numer. Methods Eng.
 \textbf{79},
1309 (2009).

\bibitem{carpet}
 J.S. Li, J.B. Pendry,
"Hiding under the Carpet: A New Strategy for Cloaking",
\newblock  Phys. Rev. Lett. \textbf{101}, 203901 (2008)

\bibitem{arbitrary}
 A. Nicolet, F. Zolla, S. Guenneau, "Electromagnetic analysis of cylindrical cloaks of an arbitrary cross section",
\newblock  Opt. Lett. \textbf{33}, 1584 (2008).

\end{thebibliography}
\end{document}